\documentclass[12pt]{iopart}

\usepackage{graphicx}

\begin{document}

\title[The measurement of scintillation emission spectra]{The measurement of scintillation emission spectra by a 
coincident photon counting technique}

\author{J E McMillan $^1$ and C J Martoff $^2$}

\address{$^1$ Department of Physics and Astronomy,
University of Sheffield, Sheffield, South Yorkshire, S3 7RH, Great Britain}
\address{$^2$ Department of Physics,
Barton Hall, Temple University, Philadelphia, PA 19122-6082, USA}

\ead{j.e.mcmillan@sheffield.ac.uk}

\begin{abstract}In the evaluation of novel scintillators, it is important to ensure that 
the spectrum of the light emitted by the scintillator is well matched to the response of
the photomultiplier.  In attempting to measure this spectrum using radioactive sources, 
it is found that so few photons are 
emitted per scintillation event
that conventional spectroscopic techniques cannot easily be used.  
A simple photon counting technique is presented, using two photomultipliers 
operated in coincidence, the one viewing the scintillator directly, 
while the other views it through a monochromator.  This system
allows the spectrum to be measured
without using specially cooled 
photomultipliers, intense radioactive sources or particle beams.   
\end{abstract}

 
\submitto{\MST}

\maketitle

\section{Introduction}

One of the most important parameters of any novel scintillator is the spectral 
response of the light produced.  For the scintillator to be 
experimentally useful, this response must
match the responses of readily available photomultiplier photocathodes 
or other photodetectors.

In the case of organic scintillators, the ultra-violet induced fluorescence 
spectrum is frequently measured, 
as this is considered to be identical to the scintillation 
spectrum \cite{Bi64b,La71b,Be71b} since both emissions are due to the same 
electronic transitions. However, it remains difficult
to study the ultra-violet induced spectrum accurately where the absorption and emission bands
of the fluorescing molecules overlap.  Further, the spectra emitted by organic 
solutions can be considerably different from the the spectra obtained by 
direct UV-excitation of the pure solutes~\cite{Be71b}.

For inorganic 
scintillators, the absorption bands are generally
of such short wavelengths that generating a suitable excitation spectrum becomes 
problematic and transmission of this through more than a few 
millimetres of material impossible~\cite{Bi64a}.  Additionally, in many 
inorganic crystals, the emission spectra 
produced are known to depend markedly on the exciting particles \cite{Va56a}. 

These considerations suggest that it would be preferable to excite
the scintillator directly with radiation, rather than to rely on the 
ultra-violet induced spectrum.  In the case of organic scintillators, 
since $\alpha$-particles will only 
penetrate the surface layers and because there are problems with absorption 
of $\gamma$-rays in small samples, previous measurements have mainly
been done using $\beta$-particles \cite{Ot56a,Bi64b,La71b}.  
However, with the $\beta$-particle energies emitted by available
radioactive sources and typical organic scintillator efficiencies
(one photon for 100eV energy deposited~\cite{Cl74a}), very little light is produced per event. 
This light must be fed through a monochromator 
and then passed to a broad band photon detector in order to derive the spectrum.  

The scintillation photons are emitted isotropically and the angular acceptance of 
monochromators is inevitably low, consequently only a tiny fraction of the 
photons produced will 
reach the detector. There is also the trade-off that if the monochromator 
is set  for higher resolution, then the throughput is unavoidably reduced 
and the signal-to-noise ratio is degraded.
Attempts have been made to avoid the problems encountered with
$\beta$-sources by using electrons from accelerators~\cite{Ho75a,Pr79a}.
While these were successful, the equipment required is unlikely to be 
generally available.
More recent work to characterize a range of inorganic compounds has made 
use of a bench-top pulsed X-ray system~\cite{Mo95a}.

The measurement of scintillation emission spectra is essentially a 
signal-to-noise ratio problem.
The signal (here, the number of photons per event) can be increased 
by depositing more energy in the scintillator.  Clearly there is a 
limit to the energy obtainable from $\beta$-particles from radioactive
sources, while the use of accelerators to produce higher energy electrons
unnecessarily complicates the measurement.  Increasing the rate
of events and integrating over many events is a workable option
but this involves working with excessively radioactive sources
and, while this may have been an option in the past, is no 
longer acceptable.

Since it is clearly difficult to increase the signal, the next obvious 
approach is to attempt to reduce the noise.
The principle source of noise is that of single-electron noise in the 
recording photomultiplier.
Cooling the photomultiplier reduces this, but since not all the noise 
is of thermal origin, this approach can only be partially successful.
The present method uses two photomultipliers and a coincidence technique 
to suppress the noise, whatever its physical origin.

\section{Previous Measurements}
Before describing the coincidence 
method in detail, it is worth reviewing the previous measurements and 
comparing their capabilities and achievements.

Ott~et~al.~\cite{Ot56a} used a photographic technique to integrate the faint 
scintillation light from organic liquid samples over a long period of time.  
A 3.7GBq $^{137}$Cs source providing $\beta$-particles of 0.51 and 1.17MeV was used.
The resolution appears to have been less than 5nm although this is not 
specified directly.
Clear inconsistencies between the fluorescence and scintillation spectra were 
observed. 

Van Sciver~\cite{Va56a} examined the spectral emission of sodium iodide
crystals, excited by either a $^{60}$Co
$\gamma$-ray source or a  $^{242}$Cm $\alpha$-particle source.  
The light was passed 
through a monochromator onto a photomultiplier cooled with dry ice
and the DC output of this was fed via an amplifier to a pen recorder.
No specification of the spectral resolution of the system is given
but responses are presented which suggest that the resolution is 5nm.

Birks~\cite{Bi64b} compared the scintillation and fluorescence spectra
of organic solutions. A grating spectrometer was coupled to a 
photomultiplier, the DC output of which was coupled to a chart recorder
to chart the spectrum. $\beta$-particles of 1.71MeV were provided by a
52MBq $^{32}$P source.
To overcome the low intensity of the light emission, the input and output 
ports of the spectrometer were widened, giving a resolution of 8nm for 
the better scintillators and 16nm for the poorer materials.

Langenscheidt \cite{La71b} further studied organic solutions,
providing a comparison of spectra obtained by
UV excitation with those obtained from $\alpha, \beta$ and 
$\gamma-$radiation.  The output of the spectrometer was coupled to a 
photomultiplier cooled to $-80^{\circ}$C, the DC output of which was 
displayed on a chart recorder.
Three radioactive sources were used, 
a 37MBq $^{241}$Am source providing 5MeV $\alpha$-particles, 
a 19MBq $^{247}$Pm source providing 225keV $\beta$-particles and
a 74GBq $^{192}$Ir source providing 0.2--1.36MeV $\gamma$-rays.
In the case of the $\beta$ source, the particle energy was so 
low that the range in the liquid was only $\sim 0.1$mm.
The resolution appears to have been around 5nm although, again, this is not 
directly specified.

Horrocks~\cite{Ho72a,Ho75a} used intense pulses of 3MeV electrons derived 
from a Van de Graaff generator.  These 
were typically 3ns long and with beam currents of 400mA.  
The test samples were organic liquid samples, flame sealed in quartz ampoules
so that they could be held in a vacuum line for 
exposure to the beam.   The light produced was passed through a 
monochromator and onto a photomultiplier, individual pulses being recorded.
A resolution of 0.7nm was attained. Problems 
were encountered both with short term quenching and with long term damage 
of the scintillator as a result of the intense radiation flux.

Pronko~et~al.~\cite{Pr79a} developed a system in which intense pulses of 
20keV electrons, produced by a simple electron gun, were used to excite 
plastic scintillator samples held in a vacuum line. 
The light was fed through a spectrometer and onto a photomultiplier.
The signal-to-noise ratio was improved by use of a lock-in amplifier
which modulated the electron beam.  
The resolution was 3nm but problems were encountered because the electron 
beam current was so high that deterioration of the light output 
of materials was observed.

Moses~et~al.~\cite{Mo95a} designed a facility for the bulk evaluation of 
novel inorganic scintillators using a pulsed X-ray tube.  Each 100ps long 
pulse produced, on average, 40 X-ray quanta per steradian with a mean 
energy of 18.5keV.  The X-rays excited crystal or powdered inorganic 
samples and the light from these was monitored by a photomultiplier.  
By placing a monochromator between the sample and the photomultiplier, 
the emission spectrum of the scintillator could be determined. A 
resolution of 12nm was achieved.  The system was limited by the low level
of light. 

Kirov~et~al.~\cite{Ki99a,Ki00a} examined the spectral emission of 
water-equivalent plastic and liquid scintillator solutions. 
A 1.7GBq $^{90}$Sr/$^{90}$Y source was used, providing $\beta$-particles of 
2.28MeV. The light from the scintillator was fed through a 
monochromator and onto a photomultiplier.  The output from this was 
integrated over a few tens of seconds and an identical time interval
dark background signal was subtracted. The monochromator accuracy 
was measured as 0.8nm though the step size used in actual 
measurements was 2.5nm.  Both the monochromator and the data 
acquisition were computer controlled allowing the automatic 
measurement of spectra.

\section{Coincident Photon Counting}
\begin{figure}[h]
 \centering
 \includegraphics{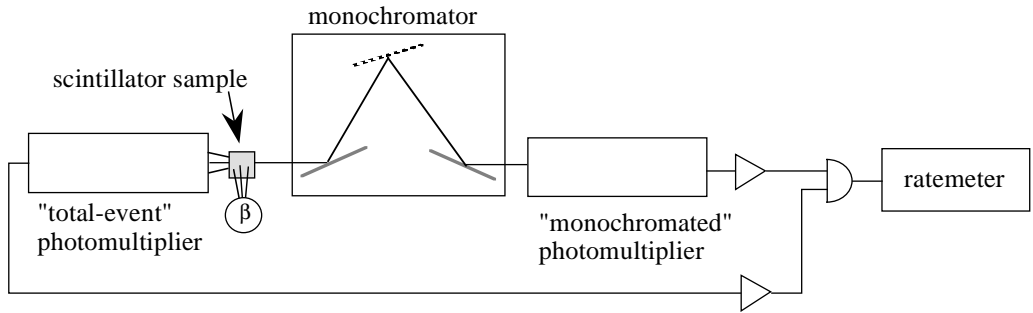}
 \caption{The experimental setup}
 \protect\label{experiment}
\end{figure}

In the coincident photon counting technique, two photomultipliers are 
used to increase the signal-to-noise ratio of the measurement.
The arrangement is shown schematically in figure~\ref{experiment}. 
One of the photomultipliers, denoted the ``total-event'' tube, 
views the scintillator sample directly, 
while the other, denoted the ``monochromated'' tube records only photons 
that have passed through the monochromator.  
The signal 
from the total-event photomultiplier is passed to a discriminator set to 
trigger at a typical $\beta$-particle flash, while the output from the monochromated 
photomultiplier is passed to a discriminator set to detect single 
photons.  The coincidence of the two discriminator outputs is taken and 
the rate of coincidences is measured in real time.   In this way, 
photons recorded by the monochromated tube are only counted if they are in 
coincidence with a scintillation flash recorded by the total-event 
photomultiplier.  This effectively suppresses the single-photon noise
of the monochromated photomultiplier.

The system was designed to study both liquid and plastic 
scintillator samples.   Liquid samples were contained in 
$10 \times 10 \times 55$mm disposable acrylic cuvettes while plastic 
samples were machined to similar dimensions.
The samples were arranged so that
$\beta$-particles entered the scintillator from below through the base.
The $\beta$-particles had an end-point energy of 2.28MeV and were obtained 
from a 3.7MBq $^{90}$Sr source. 
This source was held in a brass collimator consisting of a 50mm long
20mm diameter cylinder with a central cavity.  A 1mm 
diameter axial hole was drilled into this cavity in order to provide a well-defined 
$\beta$-particle beam with negligible exposure in other directions.
The sample was arranged so that the volume of scintillator immediately above the base was 
aligned with the input slit of the monochromator. 
\begin{figure}[h]
 \centering
\includegraphics{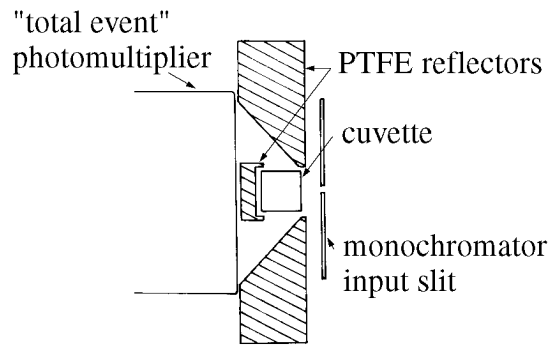}
 \caption{Plan view of the arrangement of the sample and PTFE reflectors}
 \protect\label{ptferef}
\end{figure}
The base of the acrylic cuvettes was 1mm thick, so some absorption of
$\beta$-particles in the wall could be expected. However, at 2.27MeV,
the range in material 
of density 1.0 is about 10mm so the attenuation effect is 
negligible. 
Tests with cuvettes containing only water or pure mineral oil indicated 
there was no problem with scintillation in the walls of the cuvette.
Before liquids were measured they were sparged with solvent-saturated 
nitrogen to remove dissolved oxygen.

The monochromator was an Applied Photophysics High Radiance Monochromator,
selected purely on the grounds of availability. 
Coupled to the output port of the monochromator was an 
Electron Tubes 9829QA quartz windowed photomultiplier,
chosen
to ensure that the spectral response of the measuring system 
extended into the ultraviolet.
  
A system of diffuse reflectors fabricated from PTFE 
(polytetrafluoroethylene)
ensured that, while as much light
as possible entered the input port of the monochromator,  some
of the light was directed to a second photomultiplier, the ``total-event'' 
photomultiplier,  here an Electron Tubes 9954KA.  This arrangement 
can be seen in plan view in figure~\ref{ptferef}.  For clarity, 
the radioactive 
source and its collimator are not shown; they are immediately below 
the sample cuvette. 
 
The anode signals of the photomultipliers were both fed to discriminators.
The ``total-event'' tube had an Ortec 584 with the discriminator set to 
about 100 photo-electrons while the ``monochromated'' tube had a Mechtronics
511 photon discriminator set at about 0.3 times the single photon level.
The outputs of the discriminators were set to about 50ns width and fed to 
a LeCroy 466 coincidence unit.  The coincidence output was simply counted 
over 100 second periods with a Racal 9902 counter timer and this rate 
gave a measure of the intensity of the scintillation light at the 
wavelength selected by the monochromator.  

With the resolution of the monochromator set at 5nm, and typical organic 
scintillators observed at peak wavelength, the rate of coincidences was a few 
tens per second.  This meant that statistically significant results could
be obtained with count periods of 100s per 5nm wavelength bin.  A complete
spectrum from 360nm to 540nm would take a little over an hour to complete.
The system described here was manually controlled but it would be simple
to automate the measurement by motor driving the monochromator.
It was also possible to improve the resolution at the expense of longer 
counting times.

In order to determine the relative spectral response of the recording system, tests 
were performed using light from a quartz-halogen lamp and a mercury vapour 
lamp.  The light from these was attenuated with a series of pinholes and 
fed through the monochromator and onto the quartz window of the 
photomultiplier.  For these tests, the single-photon count rate was 
determined directly without a coincidence requirement.
By varying the monochromator over the range of interest and comparing the 
single-photon rate with the known spectral emission of the sources, a set 
of correction curves for the instrumental response of the monochromator and 
quartz window photomultiplier were obtained.   
In the case of the mercury vapour lamp, individual emission lines could be
discerned, and this provided a check on the wavelength calibration of the 
monochromator.
One series of the
correction measurements was performed with a sample wall from one of the 
cuvettes inserted between the light source and the monochromator 
to ensure that the corrections included wall losses. 
  
With the 
photomultiplier used and with the organic scintillators being studied, the 
correction factors were minor.  If the technique were used 
for materials emitting further into the ultraviolet or into the green, 
where the response of the photomultiplier is reduced, the 
accuracy with which the corrections were known would begin to dominate the 
measurements. 

Since it was clear at an early stage that the coincident photon counting 
technique provided excellent signal-to-noise ratio, no effort was made to
cool the photomultipliers.  Indeed, it is not obvious that any improvement
to the measurements would be made by such cooling.

\section{Results}
\begin{figure}[h]
 \centering
 \includegraphics[width=12cm]{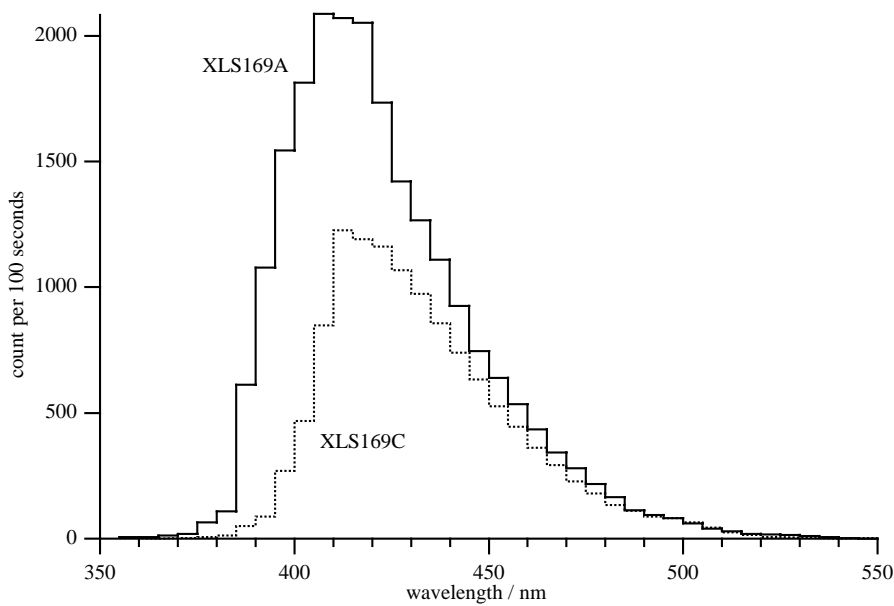}
 \caption{Typical liquid scintillator spectra}
 \protect\label{spectra}
\end{figure}

To show typical results from the system, spectra
from two liquid scintillator samples, XLS169A and 
XLS169C produced by Zinsser Analytic (UK) Ltd., are presented in 
figure~\ref{spectra}. 
Both materials are ternary scintillators using 
POPOP and bis-MSB.  
XLS169A is based on diisopropylnapthalene, whereas 
XLS169C is a mineral oil and naphthalene mixture.  It can be seen that 
while the two spectra are similar on the long wavelength side, the XLS169C cuts
off more sharply on the short wavelength side, presumably because the 
mineral oil solvent has lower transparency in the near ultraviolet.  

It is worth noting that the integral of each curve gives the total amount of 
light emitted over all wavelengths and hence provides a comparative 
measure of efficiency for the scintillator 
sample. In the work of Kirov~et~al.~\cite{Ki99a},
the corrections needed for accurate calculation of the number 
of photons per unit wavelength are considered and a detailed error 
analysis for estimating the scintillator efficiency from spectral 
measurements is given.
 
\section{Conclusions}
The coincident photon counting technique is demonstrated as a simple 
method of determining scintillator spectral response.  It requires only 
low cost equipment typically available in laboratories where 
scintillators are studied.   The apparatus described uses a relatively 
low activity $\beta$-particle source and organic scintillator samples, 
but the technique could easily 
be extended to use $\alpha$ or $\gamma$ sources and modified for inorganic
or even cryogenic scintillators.

In the case of ternary organic scintillators, employing a solute and a wavelength 
shifter, care must be taken that the light paths through scintillator are
longer than the extinction length of the wavelength shifter. If this is 
not the case, unshifted photons will escape the volume and the measured 
spectrum will be inaccurate. For a typical commercial plastic scintillator, 
Bicron BC400/NE-102, this has been measured~\cite{Qu02a} as 120$\mu$m.
With the geometry described above and materials with realistic levels of 
wavelength shifter, no problems were encountered.   

In the case of inorganic scintillators with long decay times and
more than one decay component, it might be possible  
to examine changes in spectral output as a function of time 
by using a short coincidence gate and specifying a time delay from the 
total-event discriminator.

The system could further be extended to examine the emission spectra of 
wavelength shifter materials by using a radioactive source to excite a 
scintillator which is arranged to illuminate one of the large faces 
of a rectangular slab
of polished acrylic.  The wavelength shifter can either be doped into the 
acrylic or be deposited as a film on the large face.  If light 
from the slab is coupled through one of the small edges into 
the monochromator, it will be uncontaminated by the original 
scintillation light spectrum.

The coincident photon counting technique described may also be applied to the
measurement of the spectra of any other pulsed light emission such as Cherenkov
light, plasma or spark discharges.

\ack
This work was developed as part of a Particle Physics and Astronomy Research
Council PIPSS scheme.
We would like to thank the staff of Zinsser Analytic (UK) for samples 
and assistance with this work.

\section*{References}
\bibliographystyle{unsrt}
\bibliography{spectral}

\end{document}